\documentclass[12pt]{article}
\usepackage{amsmath}
\usepackage{color}
\usepackage{graphicx}
\begin{document}
\baselineskip=18 pt
\begin{center}
{\large{\bf Aharonov-Bohm effect on spin-$0$ scalar massive charged particle with a uniform magnetic field in Som-Raychaudhuri space-time with a cosmic string }}
\end{center}

\vspace{0.5cm}

\begin{center}
{\bf Faizuddin Ahmed}\footnote{\bf faizuddinahmed15@gmail.com ; faiz4U.enter@rediffmail.com}\\
{\bf Maryam Ajmal Women's College of Science \& Technology, Hojai-782435, Assam, India}
\end{center}

\vspace{0.5cm}

\begin{abstract}

We study the relativistic quantum dynamics of spin-$0$ massive charged particle in a G\"{o}del-type space-time with electromagnetic interactions. We solve the Klein-Gordon equation subject to a uniform magnetic field in the Som-Raychaudhuri space-time with a cosmic string. In addition, we include a magnetic quantum flux into the relativistic quantum system and obtain the energy eigenvalues and analyze an analogue of the Aharonov-Bohm (AB) effect.

\end{abstract}


\vspace{0.3cm}

{\bf PACS Number(s):} 03.65.Pm, 03.65.Ge

\section{Introduction}

The relativistic quantum dynamics of spin-$0$ and spin-$\frac{1}{2}$ particles have been investigated by several reseracheres. Spin-$0$ particles such as bosons, mesons are described by Klein-Gordon equation and spin-$\frac{1}{2}$ particle such as fermions by Dirac equation. The exact solution of the wave-equations are very important since they contain all the necessary information regarding the quantum system under consideration. However, analytical solutions are possible only few cases, such as, the hydrogen atom and harmonic oscillator \cite{LL,DJG}. In recent years, many studies have been carried out to explore the relativistic energy eigenvalues and the corresponding wave-functions of these wave equations with or without external fields. The relativistic wave equations have been of current research interest for theoretical physicists \cite{BT,AWT} including in nuclear and high energy physics \cite{TYW,WG}. The relativistic quantum dynamics of spin-$0$ particles in the presence of external fields have been of great interest. The physical properties of the systems are accessed by the solution of Klein-Gordon equation with electromagnetic interactions \cite{WG,VB}. The electromagnetic interactions are introduced into the Klein-Gordon equation through the so called minimal substitution, $p_{\mu}\rightarrow p_{\mu}-e\,A_{\mu}$, where $e$ is the charge and $A_{\mu}$ is the four-vector potential of the electromagnetic field.

The relativistic quantum dynamics of spin-$0$ massive charged particles of mass $M$ is described by the KG-equation \cite{ERFM}
\begin{equation}
[\frac{1}{\sqrt{-g}}\,D_{\mu} (\sqrt{-g}\,g^{\mu\nu}\,D_{\nu})-\xi\,R-M^2]\,\Psi=0,
\label{1}
\end{equation}
where $D_{\mu}=\partial_{\mu}-i\,e\,A_{\mu}$ is the minimal substitution with $e$ is the electric charge, $A_{\mu}$ is the potential of electromagnetic field, $R$ is the scalar curvature and $\xi$ is the non-minimal coupling cosntant.

In recent years, several researchers have investigated the relativistic quantum dynamics of scalar particles in the background of G\"{o}del-type geometries. For examples, the relativistic quantum dynamics of scalar particles \cite{BDBF}, Klein-Gordon oscillator with an external fields \cite{ZW}, scalar particles with a cosmic string \cite{JC}, linear confinement of a scalar particle \cite{RLLV} (see also, \cite{EPJC}), Ground state of a bosonic massive charged particle in the presence of external fields in \cite{EOS} (see also, \cite{EPJP2}). Furthermore, the relativistic quantum dynamics of a scalar particle in the Som-Raychaudhuri metric was investigated in \cite{ND,SD} and observed the similarity of the energy eigenvalues with the Landau levels in flat space \cite{LDL,CF4}. The behavior of scalar particles with Yukawa-like confining potential in the Som-Raychaudhuri space-time in the presence of topological defects was investigated in \cite{ME}. Other works are the scalar field subject to a Cornell potential \cite{HH}, survey on the Klein-Gordon equation \cite{HH2}, bound states solution of spin-$0$ massive in a G\"{o}del-type space-time with Coulomb potential \cite{CTP}. In addition, spin-half particles have been studied in G\"{o}del-type space-time \cite{BDBF}, in the Som-Raychaudhuri space-time with torsion and cosmic string \cite{GQG}, with topological defect \cite{GQG2}, Fermi field and Dirac oscillator in the Som-Raychaudhuri space-time \cite{MM}, Dirac Fermi field with scalar and vector potentials in the Som-Raychaudhuri space-time \cite{PS}.

Our main motivation is to study the relativistic quantum dynamics of spin-$0$ scalar charged particles in the presence of an external fields including magnetic quantum flux in the Som-Raychaudhuri space-time with the cosmic string which wasn't studied in \cite{JC,ND}. We solve the Klein-Gordon equation in the considered framework and evaluate the energy eigenvalues and eigenfunctions and analyze the relativistic analogue of Aharonov-Bohm effect for bound states. We compare our results with \cite{ERFM,JC,ND} and see that the energy eigenvalues obtain here get modify due to the presence of various physical parameters.

\section{Spin-$0$ scalar massive charged particles : The KG-equation }

Consider the following Som-Raychaudhuri (SR) space-time with a cosmic-string given by \cite{JC,CTP,ME,MM,PS}
\begin{equation}
ds^2=-(dt+\alpha\,\Omega\,r^2\,d\phi)^2+\alpha^2\,r^2\,d\phi^2+dr^2+dz^2,
\label{2}
\end{equation}
where $\alpha$ and $\Omega$ characterize the cosmic string and the vorticity parameter of the space-time, respectively.
The scalar curvature $R$ of the space-time is given by
\begin{equation}
R=2\,\Omega^2.
\label{3}
\end{equation}
We choose the four-vector potential of electromagnetic fields $A_{\mu}=(0,\vec{A})$ with
\begin{equation}
\vec{A}=(0,A_{\phi},0).
\label{4}
\end{equation}  

For the geometry (\ref{2}), KG-equation (\ref{1}) becomes
\begin{eqnarray}
&&[-\frac{\partial^2}{\partial t^2}+\frac{1}{r}\frac{\partial}{\partial r}\left(r\frac{\partial}{\partial r} \right)+\left\{\frac{1}{\alpha\,r}\left (\frac{\partial}{\partial \phi}-i\, e\, A_{\phi} \right)-\Omega\, r\, \frac{\partial}{\partial  t} \right\}^2+\frac{\partial^2}{\partial z^2}\nonumber\\
&&-(M^2+2\,\xi\,\Omega^2)]\,\Psi (t,r,\phi,z)=0.
\label{5}
\end{eqnarray}
Since the line-element is independent of time and symmetrical by translations along the $z$-axis, as well by rotations. It is reasonable to write the solution to Eq. (\ref{5}) as
\begin{equation}
\Psi (t, r, \phi, z)=e^{i\,(-E\,t+l\,\phi+k\,z)}\,\psi (r),
\label{6}
\end{equation}
where $E$ is the energy of charged particle, $l=0,\pm 1,\pm 2,....$ are the eigenvalues of the $z$-component of the angular momentum operator, and $k$ are the eigenvalues of $z$-component of the linear momentum operator.

Substituting the solution (\ref{6}) into the Eq. (\ref{5}), we obtain the following equation for the radial wave-function $\psi (r)$:
\begin{equation}
\left[\frac{d^2}{dr^2}+\frac{1}{r}\frac{d}{dr}+E^2-M^2-k^2-2\,\xi\,\Omega^2-\frac{(l-e\,A_{\phi})^2}{\alpha^2 r^2}-\Omega^2 E^2 r^2-\frac{2 \Omega E}{\alpha}(l-e\,A_{\phi}) \right]\psi (r)=0.
\label{7}
\end{equation}

\subsection{Interactions with a uniform magnetic field}

Let us consider the electromagnetic four-vector potential associated with a uniform external magnetic field given by \cite{ERFM}
\begin{equation}
A_{\phi}=-\frac{1}{2}\,\alpha\,B_0\,r^2
\label{8}
\end{equation}
such that the magnetic field is along the $z$-axis $\vec{B}=\vec{\nabla}\times \vec{A}=-B_0\,\hat{k}$.

Substituting the potential (\ref{8}) into the Eq. (\ref{7}), we obtain the following radial wave equation:
\begin{equation}
\psi ''(r)+\frac{1}{r}\,\psi' (r)+\left[\lambda-\omega^2\,r^2-\frac{l^2}{\alpha^2\,r^2} \right]\,\psi (r)=0,
\label{9}
\end{equation}
where we define
\begin{eqnarray}
&&\lambda=E^2-M^2-k^2-\frac{2\,(\Omega\,E+M\,\omega_c)\,l}{\alpha}-2\,\xi\,\Omega^2,\nonumber\\
&&\omega=\sqrt{\Omega^2\,E^2+2\,M\,\omega_c\,\Omega\,E+M^2\,\omega^2_{c}}=(\Omega\,E+M\,\omega_c),\nonumber\\ \mbox{and}
&&\omega_c=\frac{e\,B_0}{2\,M}
\label{10}
\end{eqnarray}
is called the cyclotron frequency of the charged particle moving in the magnetic field.

Transforming $x=\omega\,r^2$ into the above Eq. (\ref{9}), we obtain the following differential equation
\begin{equation}
\psi ''(x)+\frac{1}{x}\,\psi' (x)+\frac{1}{x^2}\,(-\xi_1\,x^2+\xi_2\,x-\xi_3)\,\psi (x)=0,
\label{11}
\end{equation}
where
\begin{equation}
\xi_1=\frac{1}{4}\quad, \quad \xi_2=\frac{\lambda}{4\,\omega}\quad,\quad \xi_3=\frac{l^2}{4\,\alpha^2}.
\label{12}
\end{equation}

Compairing the equation (\ref{11}) with (\ref{A.1}) in appendix A, we get
\begin{eqnarray}
&&\alpha_1=1,\quad \alpha_2=0,\quad \alpha_3=0,\quad \alpha_4=0,\quad \alpha_5=0,\quad \alpha_6=\xi_1,\nonumber\\
&&\alpha_7=-\xi_2,\quad \alpha_8=\xi_3,\quad \alpha_9=\xi_1,\quad \alpha_{10}=1+2\,\sqrt{\xi_3},\nonumber\\
&&\alpha_{11}=2\,\sqrt{\xi_1},\quad \alpha_{12}=\sqrt{\xi_3},\quad \alpha_{13}=-\sqrt{\xi_1}.
\label{ss}
\end{eqnarray}

Therefore, the energy eigenvalues expression using Eqs. (\ref{12})-(\ref{ss}) into the Eq. (\ref{A.8}) in appendix A is
\begin{eqnarray}
&&E^2_{n,l}-2\,\Omega\,\left(2\,n+1+\frac{|l|}{\alpha}+\frac{l}{\alpha} \right) E_{n,l}-M^2-k^2-2\,\xi\,\Omega^2\nonumber\\
&&-2\,M\,\omega_c\,\left(2\,n+1+\frac{|l|}{\alpha}+\frac{l}{\alpha} \right)=0
\label{13}
\end{eqnarray}
with the energy eigenvalues associated with $n^{th}$ radial modes is
\begin{eqnarray}
&&E_{n,l}=\Omega\,\left(2\,n+1+\frac{l}{\alpha}+\frac{|l|}{\alpha} \right)\pm \{\Omega^2\,\left(2\,n+1+\frac{l}{\alpha}+\frac{|l|}{\alpha} \right)^2+M^2+k^2\nonumber\\
&&+2\,M\,\omega_c\,\left(2\,n+1+\frac{|l|}{\alpha}+\frac{l}{\alpha} \right)+2\,\xi\,\Omega^2 \}^{\frac{1}{2}}.
\label{14}
\end{eqnarray}
where $n=0,1,2,....$ and $k$ is a constant.

The corresponding eigenfunctions is
\begin{equation}
\psi_{n,l} (x)=|N|_{n,l}\,x^{\frac{|l|}{2\,\alpha}}\,e^{-\frac{x}{2}}\,L^{(\frac{|l|}{\alpha})}_{n} (x),
\label{15}
\end{equation} 
where $|N|_{n,l}=\left(\frac{n!}{\left(n+\frac{|l|}{\alpha}\right)!}\right)^{\frac{1}{2}}$ is the normalization constant and $L^{(\frac{|l|}{\alpha})}_{n} (x)$ is the generalized Laguerre polynomials and are orthogonal over $[0,\infty)$ with respect to the measure with weighting function $x^{\frac{|l|}{\alpha}}\,e^{-x}$ as
\begin{equation}
\int^{\infty}_{0} x^{\frac{|l|}{\alpha}}\,e^{-x} L^{(\frac{|l|}{\alpha})}_{n} L^{(\frac{|l|}{\alpha})}_{m}\,dx=\frac{\left (n+\frac{|l|}{\alpha}\right)!}{n!}\,\delta_{n m}.
\label{int}
\end{equation}

In \cite{ND}, Klein-Gordon equation in the Som-Raychaudhuri space-time without topological defects was studied. The energy eigenvalues is given by
\begin{equation}
E_{n,l}=\Omega\,(2\,n+1+l+|l|)\pm \sqrt{\Omega^2\,(2\,n+1+l+|l|)^2+M^2+k^2}.
\label{aa}
\end{equation}
Thus by comparing the result obtained in \cite{ND}, we can see that the energy eigenvalues Eq. (\ref{14}) get modify (increases) due to the presence of a uniform magnetic field $B_0$, the topological defect parameter $\alpha$, and the non-minimal coupling constant $\xi$ with the background curvature in the relativistic system.

In \cite{JC}, Klein-Gordon equation in the Som-Raychaudhuri space-time with a cosmic string was studied. The energy eigenvalues is given by
\begin{equation}
E_{n,l}=\Omega\,\left(2\,n+1+\frac{l}{\alpha}+\frac{|l|}{\alpha} \right)\pm \sqrt{\Omega^2\,\left(2\,n+1+\frac{l}{\alpha}+\frac{|l|}{\alpha} \right)^2+M^2+k^2}.
\label{16}
\end{equation}
By comparing the result without external field as obtained in \cite{JC}, we can see that the energy eigenvalues Eq. (\ref{14}) get modify (increases) due to the presence of a uniform magnetic field $B_0$ and the non-minimal coupling constant $\xi$ in the relativistic system.

In \cite{ERFM}, the relativistic quantum dynamics of a charged scalar particles in the presence of an external fields in the cosmic string space-time was studied. The energy eigenvalues is given by
\begin{equation}
E_{n,l}=\pm \sqrt{M^2+k^2+2\,M\,\omega_c\,\left(n+\frac{1}{2}+\frac{|l|}{2\,\alpha}+\frac{l}{2\,\alpha} \right)}.
\label{17}
\end{equation}
Again by comparing the energy eigenvalues Eq. (\ref{14}) with those in \cite{ERFM} or Eq. (\ref{17}) here, we can see that the present energy eigenvalues get modify due to the presence of the vorticity parameter $\Omega$ of the space-time and the non-minimal coupling constant $\xi$ with the background curvature.

\subsection{Interactions with an external field including the magnetic quantum flux}

Let us consider the system described in Eq. (\ref{7}) in the presence of an external fields in the $z$-direction. We have assumed that the topological defects ({\it e. g.}, cosmic string) has an internal magnetic flux field (with magnetic flux $\Phi_B$) \cite{YA,GAM,MSC}. The electromagnetic four-vector potential is given by the following angular component \cite{ZW,FA}:
\begin{equation}
A_{\phi}=-\frac{1}{2}\,\alpha\,B_0\,r^2+\frac{\Phi_B}{2\,\pi}.
\label{18}
\end{equation}
Here $\Phi_B=const.$ is the internal quantum magnetic flux \cite{YA,GAM,MSC} through the core of the topological defects \cite{GAM}. Three-vector potential in symmetric gauge is defined by $\vec{A}=\vec{A}_1+\vec{A}_2$ such that $\vec{\nabla}\times \vec{A}=\vec{\nabla}\times \vec{A}_1+\vec{\nabla}\times \vec{A}_2=\vec{B}=-B_0\,\hat{k}$. It is worth mentioning that this Aharonov-Bohm effect \cite{MP,VBB} has investigated in graphene \cite{RJ}, in Newtonian theory \cite{MAA}, in bound states of massive fermions \cite{VRK}, in scattering of dislocated wave fronts \cite{CC}, with torsion effects on a relativistic position-dependent mass system \cite{IJMPD,FA,F2}, bound states of spin-$0$ massive charged particles \cite{CTP,EPJP3}. In addition, this effect has investigated in the context of the Kaluza-Klein theory \cite{CF2,CF3,EVBL,EVBL2,EVBL3,AHEP2}, and with a non-minimal Lorentz-violating coupling \cite{HB}.

Substituting the potential (\ref{18}) into the Eq. (\ref{7}), we obtain the following equation
\begin{equation}
\psi ''(r)+\frac{1}{r}\,\psi' (r)+\left[\lambda_0-\omega^2\,r^2-\frac{j^2}{r^2} \right]\,\psi (r)=0,
\label{19}
\end{equation}
where
\begin{eqnarray}
&&\lambda_0=E^2-M^2-k^2-2\,(\Omega\,E+M\,\omega_c)\,j-2\,\xi\,\Omega^2,\nonumber\\
&&j=\frac{(l-\Phi)}{\alpha}.
\label{20}
\end{eqnarray}

Following the similar technique as done earlier, we obtain the relativistic eigenvalues associated with $n^{th}$ radial modes
\begin{eqnarray}
&&E_{n,l}=\Omega\,\left(2\,n+1+\frac{l-\Phi+|l-\Phi|}{\alpha} \right)\pm \{\Omega^2\,\left(2\,n+1+\frac{l-\Phi+|l-\Phi|}{\alpha} \right)^2\nonumber\\
&&+k^2+M^2+2\,m\,\omega_c\,\left(2\,n+1+\frac{l-\Phi+|l-\Phi|}{\alpha} \right)+2\,\xi\,\Omega^2 \}^{\frac{1}{2}}.
\label{23}
\end{eqnarray}
Equation (\ref{23}) is the energy spectrum of massive charged particles in the presence of an external uniform magnetic field including a magnetic quantum flux in the Som-Raychaudhuri space-time with a cosmic string. The energy eigenvalues depend on the cosmic string parameter $\alpha$, the external magnetic field $B_0$ including the magnetic quantum flux $\Phi_B$, and the non-minimal coupling constant $\xi$. We can see that the energy eigenvalues Eq. (\ref{23}) get modify in comparison to the result Eq. (\ref{14}) due to the presence of the magnetic quantum flux $\Phi_B$ which causes shifts the energy levels and gives rise to a relativistic analogue of the Aharonov-Bohm effect.

The wave-functions are given by
\begin{equation}
\psi_{n,l} (x)=|N|_{n,l}\,x^{\frac{|l-\Phi|}{2\,\alpha}}\,e^{-\frac{x}{2}}\,L^{(\frac{|l-\Phi|}{\alpha})}_{n} (x),
\label{24}
\end{equation}
$|N|_{n,l}=\left(\frac{n!}{(n+\frac{|l-\Phi|}{\alpha})!}\right)^{\frac{1}{2}}$ is the normalization constant and $L^{(\frac{|l-\Phi|}{\alpha})}_{n} (x)$ is the generalized Laguerre polynomial.

\vspace{0.5cm}
{\bf Special Case} 
\vspace{0.5cm}

We discuss a special case corresponds to zero vorticity parameter, $\Omega \rightarrow 0$. In that case, the study space-time (\ref{2}) reduces to a static cosmic string space-time.

Therefore, the radial wave-equation Eq. (\ref{19}) becomes
\begin{equation}
\psi ''(r)+\frac{1}{r}\,\psi' (r)+\left[E^2-M^2-k^2-2\,M\,\omega_c\,j-M^2\,\omega_{c}^2\,r^2-\frac{j^2}{r^2} \right]\,\psi (r)=0.
\label{25}
\end{equation}

Transforming $x=M\,\omega_c\,r^2$ into the above equation (\ref{25}), we obtain the following equation
\begin{equation}
\psi '' (x)+\frac{1}{x}\,\psi' (x)+\frac{1}{x^2}\,(-\xi_1\,x^2+\xi_2\,x-\xi_3)\psi (x)=0,
\label{31}
\end{equation}
where
\begin{equation}
\xi_1=\frac{1}{4}\quad,\quad \xi_2=\frac{E^2-M^2-k^2-2\,M\,\omega_c\,j}{4\,M\,\omega_c}\quad,\quad \xi_3=\frac{j^2}{4}.
\label{32}
\end{equation}

We obtained the following energy eigenvalues expression associated with $n^{th}$ radial modes:
\begin{eqnarray}
E_{n,l}=\pm\,\left (M^2+k^2+4\,M\,\omega_c\,\left(n+\frac{1}{2}+\frac{|l-\Phi|}{2\,\alpha}+\frac{(l-\Phi)}{2\,\alpha} \right)\right)^{\frac{1}{2}},
\label{26}
\end{eqnarray}
where $n=0,1,2,....$ and the corresponding eigenfunction is given by Eq. (\ref{24}). 

Equation (\ref{26}) is the relativistic energy eigenvalue of a massive charged particle in the presence of an external fields including a magnetic quantum flux in static cosmic string space-time. Observe that the energy eigenvalue Eq. (\ref{26}) in comparison to those result \cite{ERFM} get modify due to the presence of the magnetic quantum flux $\Phi_B$ which causes shifts the energy levels and gives rise to a relativistic analogue of the Aharonov-Bohm effect.

We can see in the above expression of the energy eigenvalues Eqs. (\ref{23}) and (\ref{26}) that the $z$-component of the angular momentum $l$ is shifted, that is,
\begin{equation}
l_{eff}=\frac{1}{\alpha}\,(l-\Phi),
\label{27}
\end{equation}
an effective angular momentum due to both the boundary condition, which states that the total angle around the string is $2\,\pi\,\alpha$, and the minimal coupling with the electromagnetic fields. We can see that the relativistic energy eigenvalues Eqs. (\ref{23}) and (\ref{26}) depend on the geometric quantum phase \cite{YA,GAM}. Thus, we have that $E_{n,l} (\Phi_B+\Phi_0)=E_{n,l \pm \tau} (\Phi_B)$, where $\Phi_0=\mp\,\frac{2\,\pi}{e}\,\tau$ with $\tau=0,1,2,..$. This dependence of the relativistic energy eigenvalues on the geometric quantum phase gives rise to a relativistic analogue of the Aharonov-Bohm effect.

Formula (\ref{24}) suggests that, when the particle circles the string, the wave-function changes according to
\begin{equation}
\Psi\rightarrow \Psi'=e^{2\,i\,\pi\,l_{eff}}\,\Psi=Exp{\{\frac{2\,\pi\,i}{\alpha}\,(l-\frac{e\,\Phi_B}{2\,\pi})\}}\,\Psi.
\label{28}
\end{equation}
An immediate consequence of Eq. (\ref{28}) is that the angular momentum operator may be redefined as
\begin{equation}
\hat{l}_{eff}=-\frac{i}{\alpha}\,(\partial_{\phi}-i\,\frac{e\,\Phi_B}{2\,\pi}),
\label{29}
\end{equation}
where the additional term, $-\frac{e\,\Phi_B}{2\,\pi\,\alpha}$, takes into account the Aharonov-Bohm magnetic flux $\Phi_B$ (internal magnetic field).

\section{Conclusions}

In this paper, we have investigated spin-$0$ massive charged particles in the presence of an external fields including a magnetic quantum flux in the Som-Raychaudhuri space-time with a cosmic string. We have introduced the electromagnetic interactions into the Klein-Gordon equation through the minimal substitution. In {\it section 2.1}, Klein-Gordon field in the background of the Som-Raychaudhuri space-time with a cosmic string in the presence of external uniform magnetic field is considered, and derived the final form of the radial wave equation. We then solved it using the Nikiforov-Uvarov method and obtained the relativistic energy eigenvalues Eq. (\ref{14}) and corresponding eigenfunctions Eq. (\ref{15}). We have seen that the relativistic energy eigenvalues depend on the cosmic string ($\alpha$), the parameter ($\Omega$) that characterise vorticity of the space-time, the external magnetic field ($B_0$), and the non-minimal coupling constant ($\xi$). We have seen the energy eigenvalues Eq. (\ref{14}) get modify (increases) in comparison to those results obtained in \cite{JC,ND} due to the presence of an external uniform magnetic field as well as the comsic string with the non-minimal coupling constant. We have also seen that the energy eigenvalues Eq. (\ref{14}) in comparison to the result in \cite{ERFM} get modify (increases) due to the presence of vorticity parameter ($\Omega$) of the space-time. In {\it section 2.2}, we have considered an external uniform magnetic field including a magnetic  quantum flux and drived the final form of the radial wave-equation. We have solved this equation using the same method and obtained the relativistic energy eigenvalues Eq. (\ref{23}) and corresponding eigenfunctions Eq. (\ref{24}). The expression for the relativistic energy eigenvalues Eqs. (\ref{23}) reveals the possibility of establishing a quantum condition between the energy eigenvalues of a massive charged particle and the parameter that characterize the vorticity of the space-time ($\Omega$). There we have discussed a special case corresponds to zero vorticity parameter and seen that the energy eigenvalues Eq. (\ref{26}) get modify (decreases) in comparison to the results in \cite{ERFM} due to the presence of a magnetic quantum flux. We have seen that the relativistic eigenvalues depend on the geometric quantum phase \cite{YA,GAM} and we have that $E_{n,l} (\Phi_B+\Phi_0)=E_{n, l \mp \tau} (\Phi_B)$, where $\Phi_0=\pm\,\frac{2\,\pi}{e}\,\tau$ with $\tau=0,1,2,..$. This dependence of the energy eigenvalues on the geometric quantum phase gives rise to an analogue of the Aharonov-Bohm effect.

In this paper, we have shown some results which are in addition to the previous results obtained in \cite{ERFM,JC,ND} present many interesting effects. This is the fundamental subject in physics and connection between these theories (gravitation and quantum mechanics) are not well understood.

\section*{Appendix A : Brief review of the Nikiforov-Uvarov (NU) method}

\setcounter{equation}{0}
\renewcommand{\theequation}{A.\arabic{equation}}

The Nikiforov-Uvarov method is helpful in order to find eigenvalues and eigenfunctions of the Schr\"{o}dinger like equation, as well as other second-order differential equations of physical interest. According to this method, the eigenfunctions of a second-order differential equation \cite{AFN}
\begin{equation}
\frac{d^2 \psi (s)}{ds^2}+\frac{(\alpha_1-\alpha_2\,s)}{s\,(1-\alpha_3\,s)}\,\frac{d \psi (s)}{ds}+\frac{(-\xi_1\,s^2+\xi_2\,s-\xi_3)}{s^2\,(1-\alpha_3\,s)^2}\,\psi (s)=0.
\label{A.1}
\end{equation}
are given by 
\begin{equation}
\psi (s)=s^{\alpha_{12}}\,(1-\alpha_3\,s)^{-\alpha_{12}-\frac{\alpha_{13}}{\alpha_3}}\,P^{(\alpha_{10}-1,\frac{\alpha_{11}}{\alpha_3}-\alpha_{10}-1)}_{n}\,(1-2\,\alpha_3\,s).
\label{A.2}
\end{equation}
And that the energy eigenvalues equation
\begin{eqnarray}
&&\alpha_2\,n-(2\,n+1)\,\alpha_5+(2\,n+1)\,(\sqrt{\alpha_9}+\alpha_3\,\sqrt{\alpha_8})+n\,(n-1)\,\alpha_3+\alpha_7\nonumber\\
&&+2\,\alpha_3\,\alpha_8+2\,\sqrt{\alpha_8\,\alpha_9}=0.
\label{A.3}
\end{eqnarray}
The parameters $\alpha_4,\ldots,\alpha_{13}$ are obatined from the six parameters $\alpha_1,\ldots,\alpha_3$ and $\xi_1,\ldots,\xi_3$ as follows:
\begin{eqnarray}
&&\alpha_4=\frac{1}{2}\,(1-\alpha_1)\quad,\quad \alpha_5=\frac{1}{2}\,(\alpha_2-2\,\alpha_3),\nonumber\\
&&\alpha_6=\alpha^2_{5}+\xi_1\quad,\quad \alpha_7=2\,\alpha_4\,\alpha_{5}-\xi_2,\nonumber\\
&&\alpha_8=\alpha^2_{4}+\xi_3\quad,\quad \alpha_9=\alpha_6+\alpha_3\,\alpha_7+\alpha^{2}_3\,\alpha_8,\nonumber\\
&&\alpha_{10}=\alpha_1+2\,\alpha_4+2\,\sqrt{\alpha_8}\quad,\quad \alpha_{11}=\alpha_2-2\,\alpha_5+2\,(\sqrt{\alpha_9}+\alpha_3\,\sqrt{\alpha_8}),\nonumber\\
&&\alpha_{12}=\alpha_4+\sqrt{\alpha_8}\quad,\quad \alpha_{13}=\alpha_5-(\sqrt{\alpha_9}+\alpha_3\,\sqrt{\alpha_8}).
\label{A.4}
\end{eqnarray}

A special case where $\alpha_3=0$, as in our case, we find
\begin{equation}
\lim_{\alpha_3\rightarrow 0} P^{(\alpha_{10}-1,\frac{\alpha_{11}}{\alpha_3}-\alpha_{10}-1)}_{n}\,(1-2\,\alpha_3\,s)=L^{\alpha_{10}-1}_{n} (\alpha_{11}\,s),
\label{A.5}
\end{equation}
and 
\begin{equation}
\lim_{\alpha_3\rightarrow 0} (1-\alpha_3\,s)^{-\alpha_{12}-\frac{\alpha_{13}}{\alpha_3}}=e^{\alpha_{13}\,s}.
\label{A.6}
\end{equation}
Therefore the wave-function from (\ref{A.2}) becomes
\begin{equation}
\psi (s)=s^{\alpha_{12}}\,e^{\alpha_{13}\,s}\,L^{\alpha_{10}-1}_{n} (\alpha_{11}\,s),
\label{A.7}
\end{equation}
where $L^{(\alpha)}_{n} (s)$ denotes the generalized Laguerre polynomial. 

The energy eigenvalues equation reduces to 
\begin{equation}
n\,\alpha_2-(2\,n+1)\,\alpha_5+(2\,n+1)\,\sqrt{\alpha_9}+\alpha_7+2\,\sqrt{\alpha_8\,\alpha_9}=0.
\label{A.8}
\end{equation}
Noted that the simple Laguerre polynomial is the special case $\alpha=0$ of the generalized Laguerre polynomila:
\begin{equation}
L^{(0)}_{n} (s)=L_{n} (s).
\label{A.9}
\end{equation}

\end{document}